\documentstyle[11pt,paspconf,epsf]{article}
\begin{document}
\begin{center}{\footnotesize to appear in: Proceedings of the International Symposium on
Astrophysics Research And Science Education at The Vatican
Observatory, June 14-21 1998, Castel Gandolfo, Italy, Ed. C. Impey}
\end{center}
\title{VLA and VLBA observations of compact radio cores in LINER galaxies}
\author{Heino Falcke}
\affil{Max-Planck-Institut f\"ur Radioastronomie, Auf dem H\"ugel 69,
D-53121 Bonn, Germany (hfalcke@mpifr-bonn.mpg.de)}
\author{Luis C. Ho}
\affil{Carnegie Observatories, 813 Santa Barbara Street, Pasadena, CA 91101,
USA (lho@ociw.edu)}
\author{James S. Ulvestad}
\affil{NRAO, P.O. Box O, 1003 Lopezville Road, Socorro, NM 87801 (julvesta@aoc.nrao.edu)}
\author{Andrew S. Wilson \& Neil M. Nagar}
\affil{Dept. of Astronomy, University of Maryland, College Park, MD
20742-2421, USA (wilson,neil@astro.umd.edu)}
\begin{abstract}
A VLA survey of nearby LINER galaxies at 15 GHz has revealed the
presence of a compact radio core in many sources. The cores seem to be
correlated with the optical activity. Follow-up VLBA observations of
the ten brightest sources confirm that these cores have brightness
temperatures $>10^8$ Kelvin, thus confirming their AGN nature. The two
brightest radio sources show extended jet-like structures and the
flat spectral indices of all cores also suggest a jet nature
rather than emission from an ADAF.
\end{abstract}
\section{Introduction}
Quite a few nearby galaxies seem to have compact radio cores in their
nuclei, the most prominent case being the Milky Way (Sgr A*). These
radio cores resemble the cores of radio-loud quasars, showing a very
high brightness temperature and a flat to inverted radio spectrum that
extends up to submm wavelengths. The size of Sgr A* is only a few
Schwarzschild radii of the central black hole (see Falcke 1996a for a
review).  Several models have been developed to explain those cores in
the context of black hole accretion: spherical accretion models (Melia
1992), advection dominated accretion flows (ADAFs, Narayan et
al.~1998), or scaled-down AGN jet models (Falcke, Mannheim, \&
Biermann 1993; Falcke \& Biermann 1996). Recent observations have
shown that the radio core in M81---in analogy to Sgr A* labelled as
M81*---is very similar to Sgr A* and also well explained by a jet
model (Falcke 1996b). This suggests that similar ultra-compact sources
can be detected in other galaxies as well.

We have therefore conducted a high-frequency survey to search for
compact, flat-spectrum radio cores in nearby galaxies and to obtain a
statistically significant sample which can help to understand the
energetic phenomena in low-luminosity active galactic nuclei (LLAGN)
and the nature of radio cores similar to Sgr A* or M81*. Here we
report first results from VLA and VLBA surveys aiming to detect such
cores.

\section{VLA survey of LINERS}
It has been known for quite a while that early-type (E and S0)
galaxies often do indeed have compact radio cores in their nuclei
(Wrobel \& Heeschen 1984; Slee et al. 1994) and that the probability
of detecting a radio core is much higher for galaxies with nuclear
optical emission-lines (O'Connell \& Dressel 1978).

Recently Ho, Filippenko \& Sargent (1995) have presented an extensive
and sensitive spectroscopical study of a magnitude-limited and
statistically well defined sample of 486 nearby elliptical and spiral
galaxies (Palomar sample). One third of the galaxies surveyed show
evidence for LINER-like activity and 13\% turned out to be Seyferts
(Ho et al.~1997). A subsample of 48 of the LINERS was recently
observed with the VLA at 5 and 8 GHz in A and B configuration (van Dyk
\& Ho 1997).  Surprisingly, almost all galaxies showed compact nuclei
at a level of at least 0.5--2 mJy; but it is not clear whether this
activity is due to a compact starburst or is AGN related. This sample
is ideally suited to search for flat-spectrum, Sgr A*-like radio cores
at high frequencies. First results of such a VLA survey were presented
in (Falcke et al. 1997). Almost half of the galaxies were indeed
detected above a $\sim5\sigma$ detection level of 1 mJy at 15 GHz and
at least a quarter of all LINERS had flat-spectrum cores. In contrast
to the steep spectrum 5 GHz emission of these galaxies (van Dyk \& Ho
1997), the flat-spectrum 15 GHz emission is well correlated with the
H$\alpha$ flux, supporting the AGN interpretation for LINERs (see
Figure 1). Moreover, the detected radio cores fall exactly on the
H$\alpha$ vs.~radio luminosity correlation predicted by the scaled
down AGN jet model (Falcke \& Biermann 1996; Falcke et al.~1997).

\begin{figure}[ht]
\plottwo{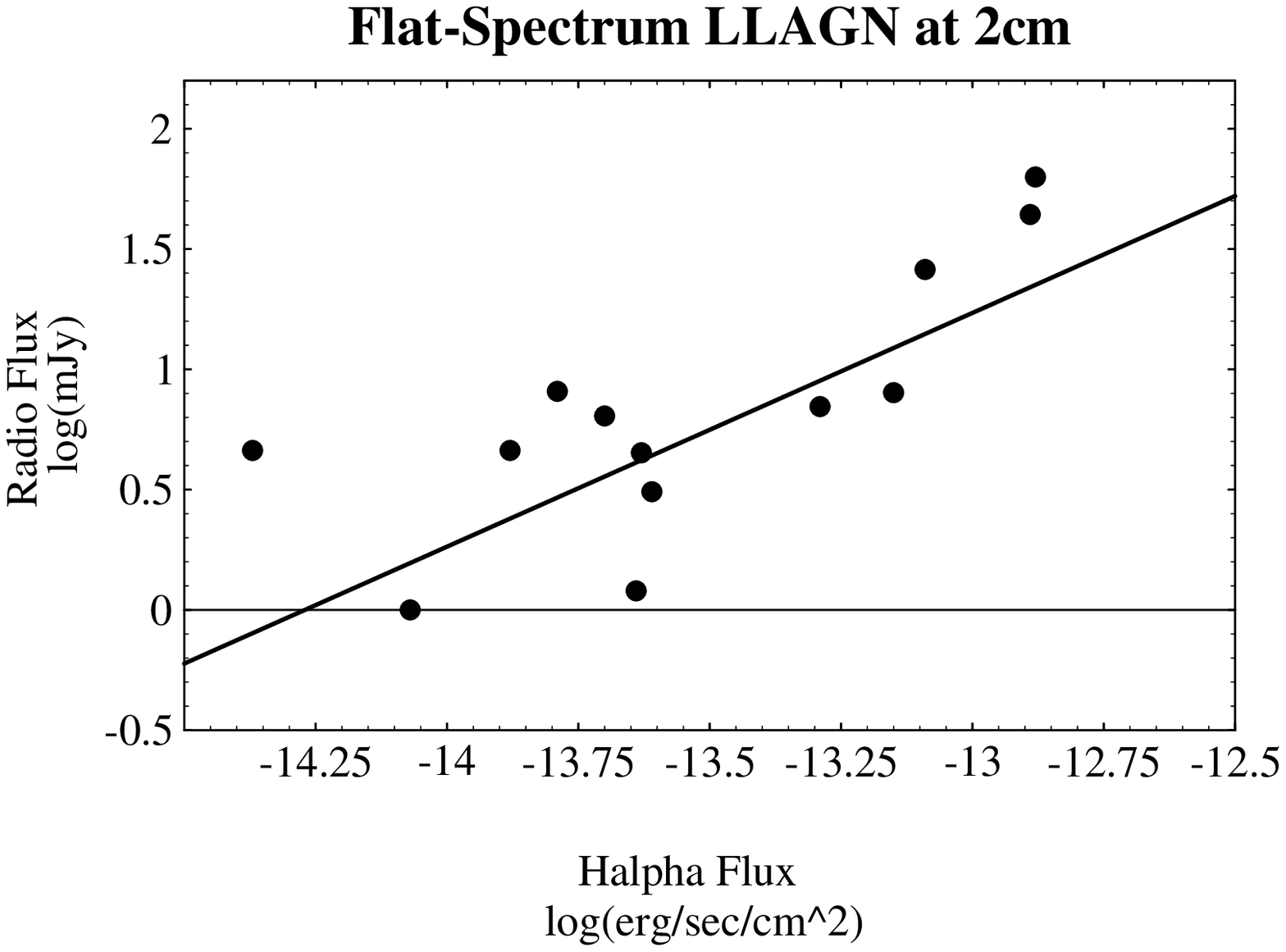}{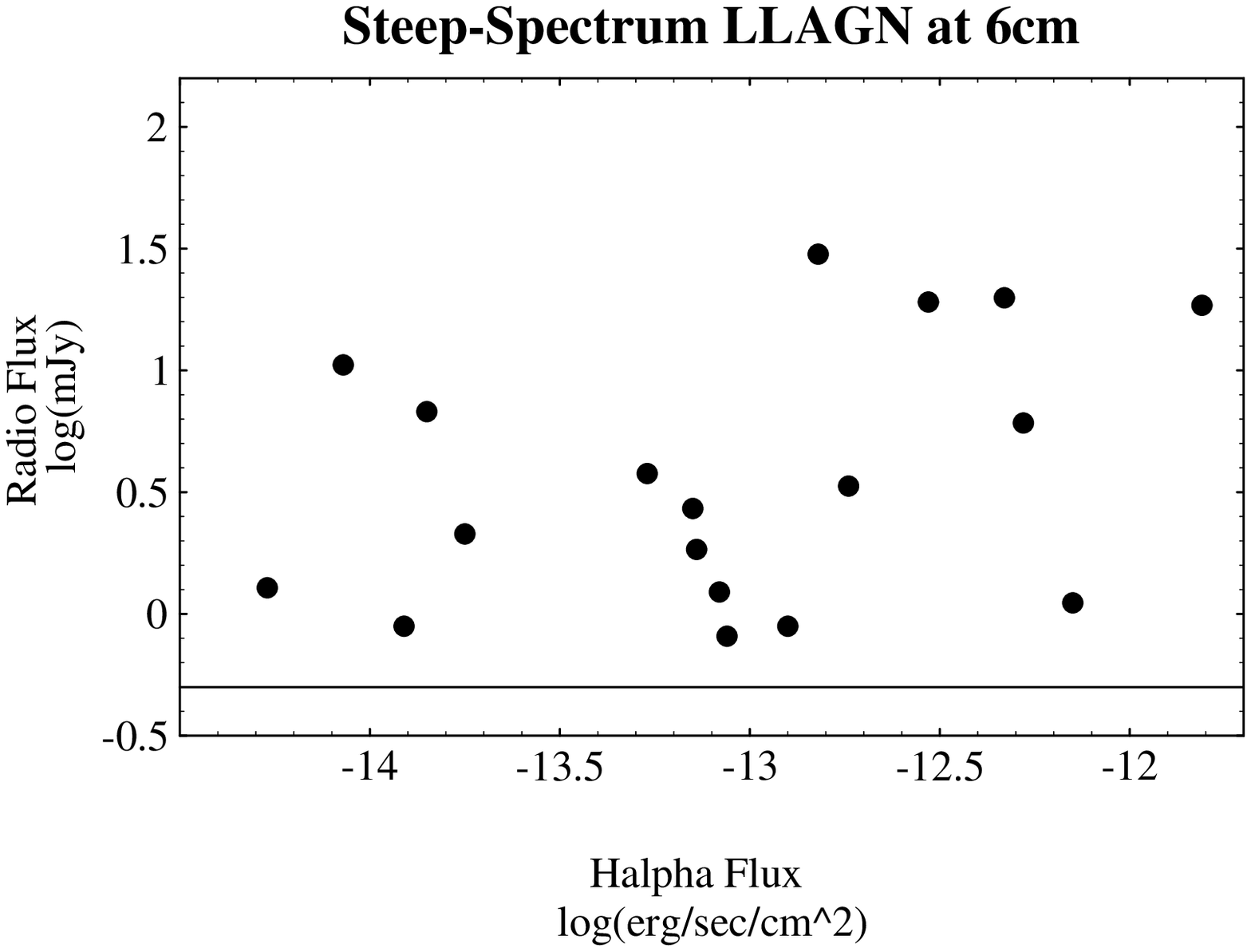}
\caption[]{\footnotesize Plots of radio flux density versus nuclear
($2^{\prime\prime}\times4^{\prime\prime}$) H$\alpha$ fluxes for a
sample of nearby LINER galaxies. Left: 15 GHz (2cm) flux density of
LINERS in the sample which were selected to have compact, flat
spectrum emission at 15 GHz; Right: 5 GHz (6cm) emission of LINERS in
the sample which were selected to have steep radio spectra. The former
are supposed to be dominated by core emission while the latter are
probably dominated by extended emission. The horizontal lines indicate
the detection limits of the surveys, the inclined solid line in the
left panel is a linear fit to the data.}
\end{figure}

\section{VLBA observation of brightest LINER cores}
In order to further test our hypothesis that the radio cores in LINERs
are indeed related to AGN activity, we have selected the eleven
brightest cores from our sample which have a flat spectrum and flux
densities larger than 3mJy. The sample was observed with the VLBA at 5
GHz in phase-referencing and snapshot mode, i.e. the telescopes
switched every few minutes from the program source to a nearby phase
calibrator source. This enabled us to detect compact
(i.e. milli-arcsecond) structure at the level of a few mJy and at
spatial scales of less than 0.1 pc.  Observations of one source were
lost because of problems with the phase-calibrator; however, all the
remaining ten sources selected from our VLA sample were indeed
detected at the flux density levels expected from extrapolating the 15
GHz VLA flux densities to 5 GHz with a flat spectrum. The two
brightest sources showed jet-like extended structure (Fig.~2), while
the remaining eight sources were basically point-like.
\begin{figure}
\plottwo{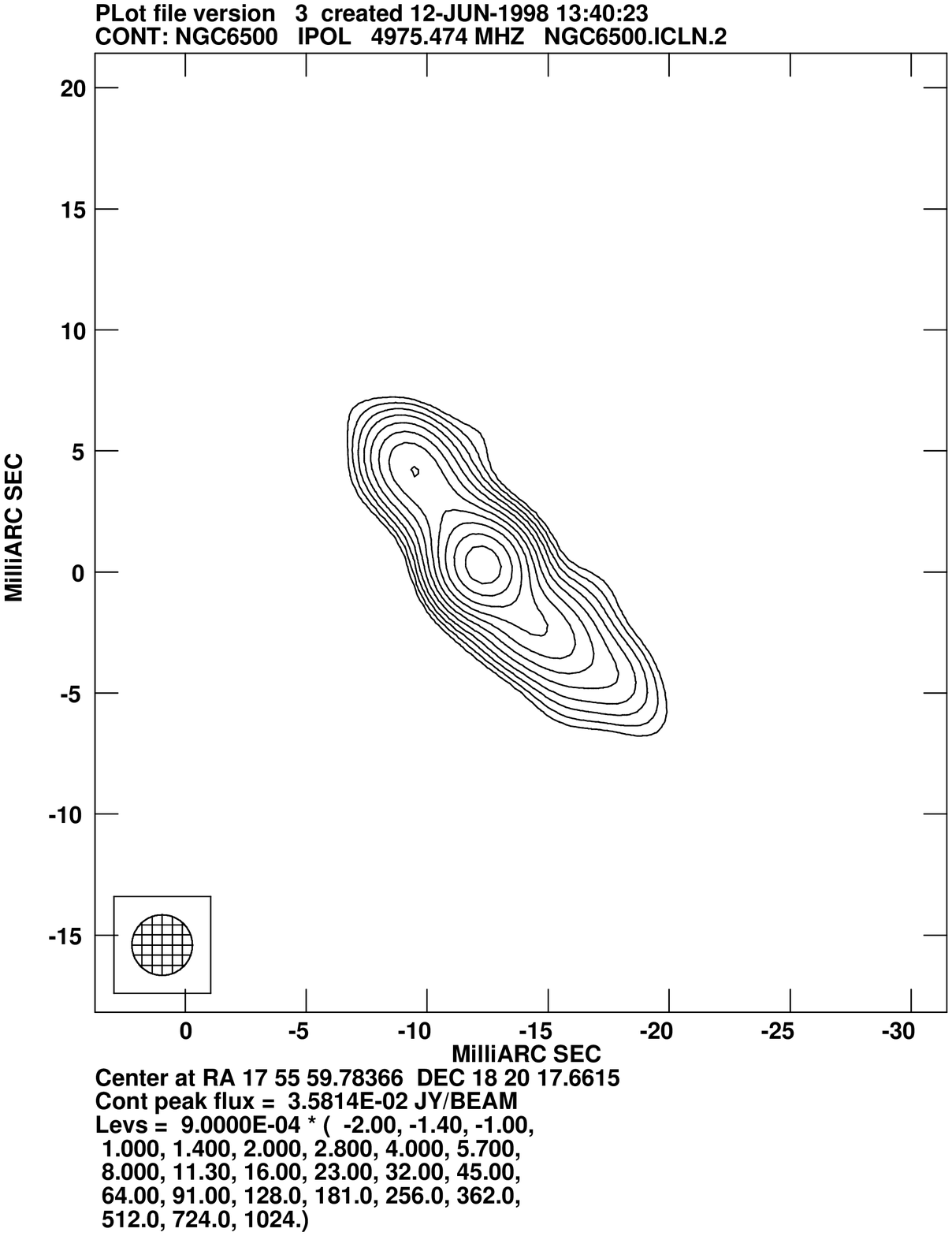}{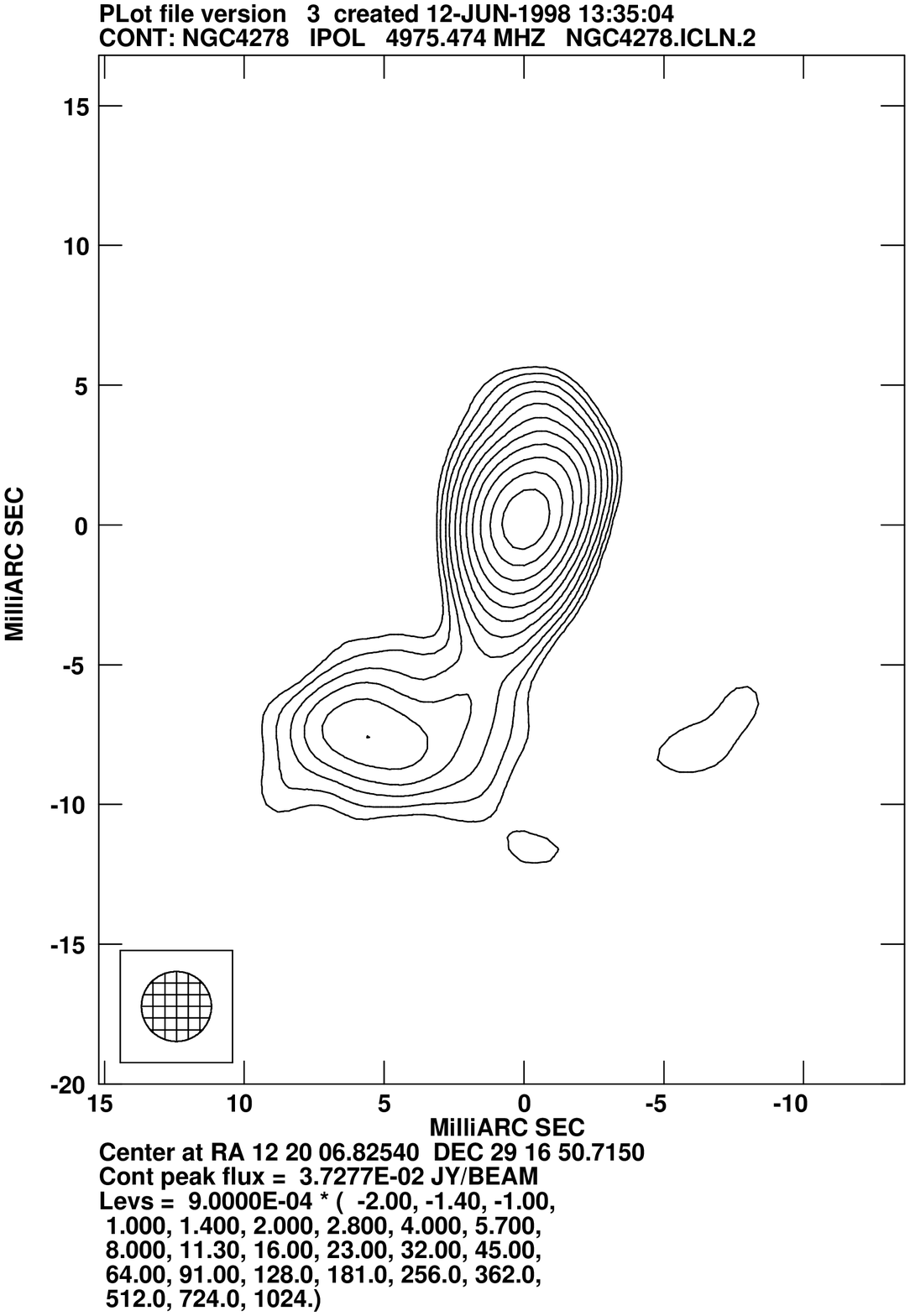}
\caption[]{\footnotesize VLBA phase-referenced and self-calibrated maps of NGC~6500
(left) and NGC~4278 (right) at 5 GHz.}
\end{figure}
This result has a number of important consequences. First of all it
demonstrates how effective our selection criterion was, giving us a
100\% detection rate despite a technically challenging project. It
seems that basically all the flux detected with the VLA is
concentrated at the milli-arcsecond scales which validates our
approach to use the VLA at 15 GHz to preferentially detect compact
radio cores.  Second, the brightness temperature of a radio source of
size 1 milli-arcsec and flux density 5 mJy at 5 GHz is T$_{b}$ $\sim$
3 $\times$ 10$^{8}$K, and indeed all our sources have lower limits for
$T_{\rm b}$ around $10^8 K$.  Hence, we can probably exclude a thermal
origin of the emission and argue that the radio emission is most
likely due to an AGN. The fact that we get similar flux densities at
15 GHz and 5 GHz also suggests a relatively flat to inverted radio
spectrum---the average 15/5 GHz spectral index of our non-simultaneous
VLBA/VLA data is $\alpha=+0.1$. This is predicted in AGN jet models
but is in stark contrast to the expectations for ADAF models, even
though we cannot exclude an ADAF component at even higher
frequencies. The idea, that these compact cores are jets is also
strengthened by the extended structures seen in the two brightest
sources.

\section{Conclusion}
Our observations have shown that galaxies with LINER-type nuclear
spectra frequently contain very compact ($<0.1$ pc) radio cores---most
likely from low power radio jets similar to those seen in Seyferts or
radio galaxies.  The monochromatic luminosities at 5 GHz of the cores
are in the range $10^{36-38}$ erg/sec and hence are comparable to M81*
but are at least four orders of magnitude more luminous than
Sgr~A*. The VLBA observations have clearly demonstrated that at least
some LINER galaxies are powered by an AGN-like engine and suggests a
continuity of AGN activity from the most luminous quasars down to
weakly active galaxies in our neighbourhood.


\begin{references}
\reference{Falcke, H. 1996a, in  ``Unsolved Problems of the
Milky Way'', IAU Symp. 169, L. Blitz \& Teuben P.J. (eds.), Kluwer, 
Dordrecht, p. 163}
\reference{Falcke, H. 1996b, ApJ, 464, L67}
\reference{Falcke, H., \& Biermann, P.L. 1996, A\&A 308, 321}
\reference{Falcke, H., Mannheim, K., \& Biermann, P. L. 1993, A\&A, 278, L1}
\reference{Falcke, H.,Wilson, A.S., Ho, L.C.  1997, in: ``Relativistic Jets
in AGN'',  Eds.  M. Ostrowski, M. Sikora, G. Madjeski, \& M. Begelman,  Cracow, p. 13}
\reference{Ho, L.C., Filippenko, A.V., \& Sargent, W.L.W. 1995, ApJS, 98, 477}
\reference{Ho, L.~C., Filippenko, A.~V., \& Sargent, W.~L.~W. 1997, ApJ, 487, 568}
\reference{Melia, F. 1992, ApJ, 387, L25}
\reference{Narayan, R., Mahadevan, R., Grindlay,  J.~E., Popham, R.G., \& Gammie, C. ~1998, ApJ 492, 554}

\reference{O'Connell, R.W., Dressel, L.L. 1978, Nature, 276, 374}
\reference{Slee, O.B., Sadler, E.M., Reynolds, J.E., Ekers, R.D. 1994, MNRAS, 269, 928}
\reference{van Dyk, S., Ho, L.C.~1997, IAU Coll. 164, A.~Zensus,
G.~Taylor, J.~Wrobel (eds.), ASP Conf.~Ser.~Vol.~144, p. 205}
\reference{Wrobel, J.M., Heeschen, D.S. 1984, ApJ, 287, 41}
\end{references}
\end{document}